\documentclass{article}%
\usepackage{amssymb}
\usepackage{amsfonts}
\usepackage{amsmath}%
\usepackage{graphicx}
\providecommand{\U}[1]{\protect\rule{.1in}{.1in}}

\begin{document}

\title{Supersymmetric approach to exact solutions of $(1+1)$-dimensional
time-independent Klein-Gordon equation : Application to a position-dependent
mass and a $\mathcal{PT}$-symmetric vector potential}
\author{N. Zaghou, F.\ Benamira and L. Guechi\\Laboratoire de Physique Th\'{e}orique, D\'{e}partement de Physique,\\Facult\'{e} des Sciences Exactes, Universit\'{e} Mentouri Constantine,\\Route d'Ain El-Bey, 25000, Constantine, Algeria}
\maketitle

\begin{abstract}
Rigorous use of SUSYQM approach applied for Klein-Gordon equation with scalar
and vector potentials is discussed. The method is applied to solve exactly,
for bound states, two models with position-dependent masses and $\mathcal{PT}%
$-symmetric vector potentials, depending on some parameters. The necessary
conditions on the parameters to get physical solutions are described. Some
special cases are also derived by adjusting the parameters of the models.


\end{abstract}

\section{Introduction}

Since the pioneering works of Bender and Boettecher \cite{Bender1,Bender2}, it
is now recognized that the hermiticity of the Hamiltonian in Schr\"{o}dinger
equation is not a necessary condition to obtain real eigenvalues for the
energy. It has been shown that, one-dimensional stationary Schr\"{o}dinger
equation may exhibit real energy eigenvalues for non-Hermitian potentials
provided that the Hamiltonian $H$ has a parity-time reversal symmetry,%
\begin{equation}
\left[  \mathcal{PT},H\right]  =0,\label{I1}%
\end{equation}
where the action of the space reflection operator $\mathcal{P}$ and the time
reversal operator $\mathcal{T}$ on position and momentum operators are given
by:
\begin{equation}
\mathcal{P}:x\rightarrow-x\text{ ; }p\rightarrow-p\text{ \ and \ }%
\mathcal{T}:x\rightarrow x\text{ ; }p\rightarrow-p\text{ ; }i\rightarrow
-i.\label{I2}%
\end{equation}

Indeed, the action of $\mathcal{PT}$ on both sides of the Schr\"{o}dinger
equation
\begin{equation}
H\psi_{n}(x)=E_{n}\psi_{n}(x),\label{I3}%
\end{equation}
combined with the properties (\ref{I1}) and (\ref{I2}) leads to
\begin{equation}
H\left[  \mathcal{PT}\psi_{n}(x)\right]  =E_{n}^{\ast}\left[  \mathcal{PT}%
\psi_{n}(x)\right]  ,\label{I4}%
\end{equation}
where $E_{n}^{\ast}$ denotes the complex conjugate of $E_{n}.$ Thus, if
$\psi_{n}(x)$ is an eigenfunction of $H$ with eigenvalue $E_{n}$, then
$\mathcal{PT}\psi_{n}(x)=\psi_{n}^{\ast}(-x)$ is also an eigenfunction of $H$
with eigenvalue $E_{n}^{\ast}$. Consequently, for eigenfunctions satisfying
$\mathcal{PT}\psi_{n}(x)=\lambda_{n}\psi_{n}(x),$ necessarily $E_{n}%
=E_{n}^{\ast}$ and vice versa, since there is no degeneracy in one dimension.
In this case, ${\mathcal{PT}}$-symmetry is said non broken, otherwise
the $\mathcal{PT}$-symmetry is said broken and the eigenvalues come in complex
conjugate pairs.\ Furthermore, since $\left(  \mathcal{PT}\right)  ^{2}=1,$
$\lambda_{n}$ are phase factors that can be absorbed in the eigenfunctions.
Hence, in the case of non broken $\mathcal{PT}$-symmetry, the eigenfunctions
may be normalized in such a way that $\mathcal{PT}\psi_{n}(x)=\psi_{n}(x).$

The normalization condition in non broken $\mathcal{PT}$-symmetric theory is
then given by \cite{Bender3,Weigert}
\begin{equation}
\int\left[  \mathcal{PT}\psi_{n}(x)\right]  \psi_{n}(x)dx=\int\psi_{n}%
^{2}(x)dx=\left(  -1\right)  ^{n}.\label{I5}%
\end{equation}

During the last two decades, solvable $\mathcal{PT}$-symmetric potentials have
been extensively studied both in relativistic and non relativistic quantum
mechanics by using different techniques
\cite{Znojil1,Znojil2,Bagchi1,Levai1,Bagchi3,Ahmed,Cannata,Simsek,Egrifes,Jia1,Levai2,Levai3,Bagchi4,Levai4,Dong,Ahmed2}%
. Moreover, some authors have investigated the solutions of Schr\"{o}dinger
equation \cite{Roy1,ZhangB,Jia2} and Dirac equation
\cite{Jia3,Jia4,Sinha3,Mustafa,Jia5,Castro} for certain $\mathcal{PT}%
$-symmetric potential models with position-dependent mass. Also, problems with
position-dependent mass in the context of Klein-Gordon and Dirac equations
with Hermitian potentials have been discussed in several works \cite{Jia6,de
Souza,Ikhdair,Dai,Ikhdair2}. However, in our knowledge, $\mathcal{PT}%
$-symmetric potentials have not been studied in the context of Klein-Gordon
equation with position-dependent mass. The aim of this work is to fill this
gap and solve exactly the $(1+1)$- dimensional time-independent Klein-Gordon
equation with position-dependent mass for bound states in the framework of
$\mathcal{PT}$-symmetry.\ In section \ref{SecII}, a summary of the approach of
supersymmetric quantum mechanics (SUSYQM) \cite{Cooper4,Cooper5} is outlined
for $\mathcal{PT}$-symmetric potentials. In section \ref{SecIII}, we show how
to map Klein-Gordon equation for position-dependent mass with mixing scalar
and vector potentials into a Schr\"{o}dinger-like equation with constant mass
and energy-dependent effective potential, suitable for processing by the
SUSYQM approach.\ Section \ref{SecIV} is devoted to applications, where we
solve exactly two problems with suitable mass distribution-functions in the
presence of $\mathcal{PT}$-symmetric vector potentials and null scalar
potentials, by the approach of SUSYQM. \ \ 

\section{Basic concepts of SUSYQM approach with $\mathcal{PT}$-symmetric
Hamiltonian \label{SecII}}

In connection to the formalism of SUSYQM for Hermitian Hamiltonians
\cite{Cooper4,Cooper5}, bound-state eigenvalues and corresponding
eigenfunctions of a $\mathcal{PT}$-symmetric one-dimensional Hamiltonian,
$H=-\frac{\hbar^{2}}{2m}\frac{d^{2}}{dx^{2}}+V(x),$ with non broken
$\mathcal{PT}$-symmetry (real eigenvalues), may be obtained in the same
way.\ The partner Hamiltonians $H^{\left(  -\right)  }$and $H^{\left(
+\right)  }$ associated to $H$ are defined as
\begin{equation}
H^{\left(  -\right)  }=H-E_{0}=BA,\label{14a}%
\end{equation}
and
\begin{equation}
H^{\left(  +\right)  }=AB,\label{14b}%
\end{equation}
where $E_{0}$ is the ground-state energy of the Hamiltonian $H,$ with
\begin{equation}
A=\frac{\hbar}{\sqrt{2m}}\frac{d}{dx}+W(x)\text{,\ \ \ \ \ \ }B=-\frac{\hbar
}{\sqrt{2m}}\frac{d}{dx}+W(x),\label{14c}%
\end{equation}
and the superpotential $W(x)$ is a complex function.

Hence, according to Eq. (\ref{14a}), $H$ and $H^{\left(  -\right)  }$ have the
same eigenfunctions $\left(  \psi_{n}\left(  x\right)  \sim\psi_{n}^{\left(
-\right)  }\left(  x\right)  \right)  $ and the eigenvalues $\left\{
E_{n}^{\left(  -\right)  }=E_{n}-E_{0}\right\}  $ corresponding to $H^{\left(
-\right)  }$ are semi-positive definite:
\begin{equation}
H^{\left(  -\right)  }\psi_{n}^{\left(  -\right)  }\left(  x\right)
=BA\psi_{n}^{\left(  -\right)  }\left(  x\right)  =E_{n}^{\left(  -\right)
}\psi_{n}^{\left(  -\right)  }\left(  x\right)  ,\label{a4}%
\end{equation}
with%
\begin{equation}
E_{0}^{\left(  -\right)  }=0\text{ and }E_{n}^{\left(  -\right)  }>0\text{ for
}n=1,2,\cdots.\label{a5}%
\end{equation}

Assuming that the ground-state eigenfunction $\psi_{0}^{\left(  -\right)
}\left(  x\right)  $ satisfies $A\psi_{0}^{\left(  -\right)  }\left(
x\right)  =0,$ it is given by%
\begin{equation}
\psi_{0}^{\left(  -\right)  }\left(  x\right)  =N_{0}\exp\left(  -\frac
{\sqrt{2m}}{\hbar}\int^{x}W(y)dy\right)  ,\label{a5a}%
\end{equation}
where $N_{0}$ is a normalization constant such that $\psi_{0}^{\left(
-\right)  }\left(  x\right)  $ is square integrable in the sense of (\ref{I5}).

The action of the operator $A$ on both sides of Eq. (\ref{a4}) leads to%
\begin{equation}
H^{\left(  +\right)  }\left(  A\psi_{n}^{\left(  -\right)  }\left(  x\right)
\right)  =E_{n}^{\left(  -\right)  }\left(  A\psi_{n}^{\left(  -\right)
}\left(  x\right)  \right)  ,\label{a5b}%
\end{equation}
such that the eigenvalues $E_{n}^{\left(  +\right)  }$ of $H^{\left(
+\right)  } $ and the normalized corresponding eigenfunctions $\psi
_{n}^{\left(  +\right)  }\left(  x\right)  $ for $n=0,1,2,\cdots,$ are related
to those of $H^{\left(  -\right)  }$ by \cite{Cooper4,Cooper5}%
\begin{equation}
E_{n}^{\left(  +\right)  }=E_{n+1}^{\left(  -\right)  }\text{ },\label{a5c}%
\end{equation}
and%
\begin{equation}
\psi_{n}^{\left(  +\right)  }\left(  x\right)  =\frac{1}{\sqrt{E_{n+1}%
^{\left(  -\right)  }}}A\psi_{n}^{\left(  -\right)  }\left(  x\right)
.\label{a5d}%
\end{equation}

Explicitly, the partner Hamiltonians reads%
\begin{equation}
H^{\left(  \mp\right)  }=-\frac{\hbar^{2}}{2m}\frac{d^{2}}{dx^{2}}+V^{\left(
\mp\right)  }(x),\label{16a}%
\end{equation}
where the partner potentials, $V^{\left(  \mp\right)  }(x)$, are given by
\begin{equation}
V^{\left(  \mp\right)  }(x)=W^{2}(x)\mp\frac{\hbar}{\sqrt{2m}}\frac{dW(x)}%
{dx}.\label{17a}%
\end{equation}

The partner potentials are said shape-invariant potentials \cite{Gendenshtein}
if they satisfy
\begin{equation}
V^{\left(  +\right)  }(x;\left\{  a_{1}\right\}  )=V^{\left(  -\right)
}(x;\left\{  a_{2}\right\}  )+R\left(  \left\{  a_{1}\right\}  \right)
,\label{a6}%
\end{equation}
where$\left\{  a_{1}\right\}  $ and $\left\{  a_{2}\right\}  $ are two sets of
real parameters related by a certain function $\left(  \left\{  a_{2}\right\}
=f\left(  \left\{  a_{1}\right\}  \right)  \right)  $ and the remainder
$R\left(  \left\{  a_{1}\right\}  \right)  $ is independent of $x.$

If the requirement (\ref{a6}) is satisfied, one can show
\cite{Cooper4,Cooper5} that the energy spectrum of $H^{\left(  -\right)  }$
can be deduced algebraically and is given by%
\begin{equation}
E_{0}^{\left(  -\right)  }=0\text{ , \ \ \ }E_{n}^{\left(  -\right)  }%
=\sum_{k=1}^{n}R\left(  \left\{  a_{k}\right\}  \right)  \text{ for
}n=1,2,\cdots,\label{a7}%
\end{equation}
with$\left\{  a_{k}\right\}  =\underset{\left(  k-1\right)  \text{
times}}{\underbrace{f\circ f\circ\ldots\circ f}}\left(  \left\{
a_{1}\right\}  \right)  .$

The spectrum of the Hamiltonian $H$ \ is then given by
\begin{equation}
E_{n}=E_{n}^{\left(  -\right)  }+E_{0}.\label{A4}%
\end{equation}

The unnormalized eigenfunctions of the excited states are given by the
recurrence formula \cite{Cooper4,Cooper5,Dutt,Dabrowska}%
\begin{equation}
\psi_{n}\left(  x;\left\{  a_{1}\right\}  \right)  =B\left(  \left\{
a_{1}\right\}  \right)  \psi_{n-1}\left(  x;\left\{  a_{2}\right\}  \right)
,\text{ for }n\geq1,\label{a8}%
\end{equation}
which leads to the general formula
\begin{equation}
\psi_{n}\left(  x;\left\{  a_{1}\right\}  \right)  =\left[  \prod_{i=1}%
^{n}B\left(  \left\{  a_{i}\right\}  \right)  \right]  \psi_{0}\left(
x,\left\{  a_{n+1}\right\}  \right)  \text{ for }n\geq1.\label{A17}%
\end{equation}

\section{$\left(  1+1\right)  $- Dimensional time-independent Klein-Gordon
equation with position-dependent mass and mixing scalar and vector potentials
\label{SecIII}}

The one-dimensional time-independent Klein-Gordon equation for a spinless
particle with position-dependent mass $M(x),$subjected to mixing vector and
scalar potentials $V(x)$ and $S\left(  x\right)  $, reads ($\hbar$ is the
Plank constant, $c$ is the speed of light)%
\begin{equation}
\left(  -\hbar^{2}\frac{d^{2}}{dx^{2}}+\frac{1}{c^{2}}\left[  \left(
M(x)c^{2}+S\left(  x\right)  \right)  ^{2}-\left(  E-V\left(  x\right)
\right)  ^{2}\right]  \right)  \varphi\left(  x\right)  =0,\label{1}%
\end{equation}
where $E$ is the energy of the particle and $\varphi\left(  x\right)  $ its
corresponding wavefunction. Since Eq. (\ref{1}) is not an eigenvalues
equation, like Schr\"{o}dinger equation, it is not easy to use SUSYQM approach
to solve it and obtain the energy spectrum algebraically.\ To overcome this
difficulty, equation (\ref{1}) is often written, in the literature, as an
eigenvalues equation in the form%
\begin{equation}
\left(  -\hbar^{2}\frac{d^{2}}{dx^{2}}+V_{eff}(x)\right)  \varphi\left(
x\right)  =\widetilde{E}\varphi\left(  x\right)  ,\label{1-1}%
\end{equation}
with
\begin{equation}
V_{eff}(x)=\frac{1}{c^{2}}\left[  \left(  M(x)c^{2}+S\left(  x\right)
\right)  ^{2}-V^{2}\left(  x\right)  +2EV\left(  x\right)  \right]
,\label{1-2}%
\end{equation}
and
\begin{equation}
\widetilde{E}=\frac{E^{2}}{c^{2}}.\label{1-3}%
\end{equation}

The disadvantage in doing so is that $E$ appears in both the effective
potential $V_{eff}(x)$ and the eigenvalue $\widetilde{E}$. When using SUSYQM
approach, the energy $E$ in the hierarchical partner potentials is considered
as a parameter that remains unchanged but it changes in the hierarchical
eigenvalues, which leads to confusion. In this work, we will follow a
different approach that removes the ambiguity.

First, remark that equation (\ref{1}) may be seen as a zero energy
Schr\"{o}dinger-like equation for a particle with constant mass ($m=1/2$),
subjected to the $E$-dependent potential%

\begin{equation}
V_{E}(x)=\frac{1}{c^{2}}\left[  \left(  M(x)c^{2}+S\left(  x\right)  \right)
^{2}-\left(  E-V\left(  x\right)  \right)  ^{2}\right]  .\label{2}%
\end{equation}

Thus, the problem reduces to search the solution of a zero-energy
Schr\"{o}dinger-like equation with a conditional parameter in potential. To
solve a typical equation for bound states, i.e. discrete real energies
$E\equiv E_{n} $ and normalized wavefunctions $\varphi\left(  x\right)
\equiv\varphi_{n}\left(  x\right)  $, using SUSYQM, we consider instead the
Schr\"{o}dinger equation $\left(  2m=1\right)  $%
\begin{equation}
\left(  -\hbar^{2}\frac{d^{2}}{dx^{2}}+V_{E}(x)\right)  \Phi_{n}\left(
x\right)  =\epsilon_{n}\Phi_{n}\left(  x\right)  \text{ \ with }%
n=0,1,2,\ldots,\label{3}%
\end{equation}
where $E$ \ is considered as a real parameter in the potential $V_{E}%
(x)$.\ Now, SUSYQM approach can be applied to solve equation (\ref{3}) without
any confusion.\ When $V_{E}(x)$ is Hermitian or $\mathcal{PT}$-symmetric and
the $\mathcal{PT}$-symmetry is not spontaneously broken, the eigenvalues
$\epsilon_{n}$ are real functions of the parameter $E.$ Hence, once the
eigenvalues $\epsilon_{n}$ and the corresponding eigenfunctions $\Phi
_{n}\left(  x\right)  $ for Eq. (\ref{3})\ are obtained, the energies $E_{n}$
of the original problem (equation (\ref{1})) are given by the real solutions
of the equation%
\begin{equation}
\epsilon_{n}\left(  E\right)  =0,\label{4}%
\end{equation}
and the wavefunctions $\psi_{n}\left(  x\right)  $ can be deduced by%
\begin{equation}
\varphi_{n}\left(  x\right)  =\left.  \Phi_{n}\left(  x\right)  \right\vert
_{\epsilon_{n}\left(  E\right)  =0}.\label{5}%
\end{equation}

\section{Applications \label{SecIV}}

We are interested in this paper to solve exactly Eq. (\ref{3}) for two models
with $\mathcal{PT}$-symmetric $E$-dependent potential.\ resulting from complex
$\mathcal{PT}$-symmetric vector potential and null scalar potential. For each
model, the mass distribution is suitably chosen such that to obtain an
$E-$dependent potential that is exactly solvable for bound states.

\subsection{Model with asymptotically unbounded mass, coupled to a linear
$\mathcal{PT}$-symmetric vector potential}

Consider a relativistic position-dependent spinless particle moving on the
whole $X$-axis and subjected to a linear $\mathcal{PT}$-symmetric vector
potential and null scalar potential. The mass distribution and the vector
potential are taken, respectively, as
\begin{equation}
M\left(  x\right)  =\sqrt{\mu^{2}+\left(  \frac{\lambda}{c}\right)  ^{2}x^{2}%
},\label{8}%
\end{equation}
and%
\begin{equation}
V\left(  x\right)  =ic\eta x,\label{9}%
\end{equation}
where $\mu$ is the value of the mass at the origin of the coordinate,
$\lambda$ and $\eta$ are real parameters with dimension $MT^{-1},$ and without
loss of generality $\lambda$ is assumed to be positive.$\ $Note that the speed
of light $c$ is explicitly included in the expressions of $M\left(  x\right)
$ and $V\left(  x\right)  $ only by convenience of calculations. However,
$\lambda$ and $\eta$ may be seen of order $0$ and order $1$ compared to
$c^{-1}$ respectively, i.e.,
\begin{equation}
\lambda=\lambda_{0}+O\left(  c^{-1}\right)  \text{ and }\eta=\eta_{0}%
c^{-1}+O\left(  c^{-2}\right)  .\label{9-1}%
\end{equation}

Substituting Eqs (\ref{8}) and (\ref{9}) into (\ref{2}), the $E$-dependent
potential reads
\begin{equation}
V_{E}\left(  x\right)  =\left(  \lambda^{2}+\eta^{2}\right)  x^{2}%
+i\frac{2\eta E}{c}x+\frac{\mu^{2}c^{4}-E^{2}}{c^{2}},\label{10}%
\end{equation}
which is a $\mathcal{PT}$-symmetric function $\left(  V_{E}\left(  x\right)
=V_{E}^{\ast}\left(  -x\right)  \right)  $ for real values of the energy $E.$

To solve Eq. (\ref{3}) with the potential (\ref{10}), by using SUSYQM, we
choose the superpotential in the form%
\begin{equation}
W_{E}(x)=Ax+i\frac{\eta E}{cA},\label{11}%
\end{equation}
where $A$ is a real parameter.\ 

In order to fix the parameter $A$ and obtain the ground-state eigenvalue
$\epsilon_{0}\equiv\epsilon_{0}\left(  E\right)  $, we have to solve the identity%

\begin{equation}
V_{E}\left(  x\right)  -\epsilon_{0}=W_{E}^{2}(x)-\hbar W_{E}^{\prime
}(x).\label{12}%
\end{equation}
\qquad

Substituting (\ref{10}) and (\ref{11}) into (\ref{12}) and identifying the
coefficients of terms in power of $x$, leads to%
\begin{equation}
A^{2}=\lambda^{2}+\eta^{2},\label{13}%
\end{equation}
and%
\begin{equation}
\epsilon_{0}=\mu^{2}c^{2}+\hbar A-\frac{\lambda^{2}E^{2}}{c^{2}A^{2}%
}.\label{15}%
\end{equation}

Note that $\epsilon_{0}$ is real for real values of $E.$ \ However, using Eq.
(\ref{a5a}), the unnormalized ground-state eigenfunction may be put in the form%

\begin{equation}
\Phi_{0}\left(  x\right)  \sim e^{-\frac{A}{2\hbar}\left(  x+i\frac{\eta
E}{cA^{2}}\right)  ^{2}}.\label{16}%
\end{equation}

By demanding that $\Psi_{0}\left(  x\right)  $ is normalizable on the real
axis in the sense of Eq. (\ref{I5}) requires that $A$ is positive, such that
the acceptable solution of (\ref{13}) is%
\begin{equation}
A=\sqrt{\lambda^{2}+\eta^{2}}.\label{17}%
\end{equation}

The supersymmetric partner potentials $V_{E}^{\left(  \mp\right)  }\left(
x\right)  =W_{E}^{2}(x)\mp\hbar W_{E}^{\prime}(x)$ are explicitly given by%

\begin{align}
V_{E}^{\left(  -\right)  }\left(  x\right)   & =A^{2}x^{2}+i\frac{2\eta E}%
{c}x-\frac{\eta^{2}E^{2}}{c^{2}A^{2}}-\hbar A,\label{18}\\
V_{E}^{\left(  +\right)  }\left(  x\right)   & =A^{2}x^{2}+i\frac{2\eta E}%
{c}x-\frac{\eta^{2}E^{2}}{c^{2}A^{2}}+\hbar A.\label{19}%
\end{align}

They satisfy the shape invariance condition (\ref{a6}), which reads%
\begin{equation}
V_{E}^{\left(  +\right)  }\left(  x,a_{1}\right)  =V_{E}^{\left(  -\right)
}\left(  x,a_{2}\right)  +R\left(  a_{1}\right)  ,\label{20}%
\end{equation}
with
\begin{equation}
a_{1}=A,\text{ }a_{2}=f\left(  a_{1}\right)  =a_{1},\label{20-1}%
\end{equation}
and
\begin{equation}
R\left(  a_{1}\right)  =2\hbar a_{1}=2\hbar A.\label{20-2}%
\end{equation}

Using (\ref{A4}), the energy eigenvalues corresponding to the potential
$V_{E}(x)$ are given by%
\begin{equation}
\epsilon_{n}=\epsilon_{n}^{\left(  -\right)  }+\epsilon_{0},\label{21}%
\end{equation}
where $\epsilon_{n}^{\left(  -\right)  }$ are the eigenvalues of the partner
$V_{E}^{\left(  -\right)  }\left(  x\right)  ,$ which are expressed in terms
of the remainder function $R$ as%
\begin{equation}
\epsilon_{n}^{\left(  -\right)  }=\sum_{k=1}^{n}R\left(  a_{k}\right)
=2n\hbar A,\label{21-0}%
\end{equation}
and we have used the fact that
\begin{equation}
a_{k}=\underset{\left(  k-1\right)  \text{ }\mathrm{times}}{\underbrace{f\circ
f\circ\cdots\circ f}}\left(  a_{1}\right)  \text{ }=a_{1}.\label{21-1}%
\end{equation}

Since $a_{k}=A>0$ for all $k=1,2,\cdots,$ the number of eigenvalues is
unlimited. Thus, substituting (\ref{15}) and (\ref{21-0}) into (\ref{21}), the
energy eigenvalues $\epsilon_{n}$ are given by
\begin{equation}
\epsilon_{n}=\mu^{2}c^{2}+\left(  2n+1\right)  \hbar A-\frac{\lambda^{2}E^{2}%
}{c^{2}A^{2}}.\label{22}%
\end{equation}

By virtue of (\ref{4}), the generating formula for allowed energy values of
the original problem, $E_{n}$, may be put in the form%
\begin{equation}
E_{n}=\pm\frac{A}{\lambda}\sqrt{\mu^{2}c^{4}+\left(  2n+1\right)  \hbar
c^{2}A},\text{ \ \ \ }n=0,1,2,\cdots.\label{23}%
\end{equation}

Thus, all the energy values are real, independently of the parameters
\ $\mu,\lambda$ and $\eta,$ and consequently the $\mathcal{PT}$-symmetry is
always not broken.\ 

To determine the wavefunctions $\psi_{n}\left(  x\right)  $ of the original
problem $\left(  \psi_{n}\left(  x\right)  =\left.  \Phi_{n}\left(  x\right)
\right\vert _{\epsilon_{n}\left(  E\right)  =0}\right)  $, let us write Eq.
(\ref{3}) for $\epsilon_{n}=0$ in the form%
\begin{equation}
\left(  -\hbar^{2}\frac{d^{2}}{dx^{2}}+A^{2}\left(  x+i\frac{\eta E_{n}%
}{cA^{2}}\right)  ^{2}\right)  \psi_{n}\left(  x\right)  =\left(  2n+1\right)
\hbar A\psi_{n}\left(  x\right)  ,\label{24}%
\end{equation}
where we made use of (\ref{5}), (\ref{10}) and (\ref{22}).

By defining a new function $\varphi_{n}\left(  z\right)  $ by
\begin{equation}
\psi_{n}\left(  x\right)  =e^{-\frac{Z^{2}}{2}}\varphi_{n}\left(  z\right)
,\label{25}%
\end{equation}
with $z=\sqrt{\frac{A}{\hbar}}\left(  x+i\frac{\eta E_{n}}{cA^{2}}\right)  $
and substituting into (\ref{24}), it is easily seen that $\varphi_{n}(z)$
satisfies the Hermit equation%
\begin{equation}
\varphi_{n}^{\prime\prime}\left(  z\right)  -2z\varphi_{n}^{\prime}\left(
z\right)  +2n\varphi_{n}\left(  z\right)  =0.\label{26}%
\end{equation}

Hence, the wavefunctions may be written in the form%
\begin{equation}
\psi_{n}\left(  x\right)  =\left\vert N_{n}\right\vert e^{i\frac{n\pi}{2}%
}e^{-\frac{A}{2\hbar}\left(  x+i\frac{\eta E_{n}}{cA^{2}}\right)  ^{2}}%
H_{n}\left(  \sqrt{\frac{A}{\hbar}}\left(  x+i\frac{\eta E_{n}}{cA^{2}%
}\right)  \right)  ,\label{27}%
\end{equation}
where $H_{n}\left(  z\right)  $ is the Hermit polynomial and $\left\vert
N_{n}\right\vert $ is a real normalizing factor. The phase factor
$e^{i\frac{n\pi}{2}}$ is introduced explicitly in order to make the
wavefunction $\psi_{n}\left(  x\right)  $ also eigenfunction of the
$\mathcal{PT}$ operator with eigenvalue equal to $1$.

Normalizing $\psi_{n}\left(  x\right)  $ in the sense of Eq. (\ref{I5}) allows
to fix the normalization factor in the form%
\begin{equation}
\left\vert N_{n}\right\vert =\left(  \frac{A}{\pi\hbar}\right)  ^{\frac{1}{4}%
}\left(  \frac{1}{2^{n}n!}\right)  ^{\frac{1}{2}}.\label{29}%
\end{equation}

In conclusion, it is obvious from Eq. (\ref{23}) that we have to consider
$\lambda\neq0$ with arbitrary $\eta$. This means that the position dependence
of the mass which is responsible of the existence of bound states.\ The vector
potential only contributes to the magnification of the energy values.\ In
other words, for a fixed value of $\lambda,$ the energy values are amplified
with increasing values of $\left\vert \eta\right\vert . $ However, in the
Hermitian version of the problem, with $V(x)=\eta cx$, the parameter $\eta$ is
to be replaced by $-i\eta$ in Eq.\ (\ref{23}) so that the role of the vector
potential is inverted. Indeed, in this case, for a fixed value $\lambda$,
energy values decrease with increasing values of $\left\vert \eta\right\vert $
and bound states exist only if $\left\vert \eta\right\vert <\lambda.$

\subsubsection{Special cases \label{SC1}}

\begin{itemize}
\item Setting $\mu=0$ in (\ref{8}), the problem reduces to a particle with a
mass distribution as a linear function of the position, given by%
\begin{equation}
M\left(  x\right)  =\frac{\lambda}{c}\left\vert x\right\vert ,\label{30}%
\end{equation}
subjected to the $\mathcal{PT}$-symmetric vector potential (\ref{9}). This
special case may also be seen as the problem of massless particle subjected to
the $\mathcal{PT}$-symmetric vector potential (\ref{9}), combined with a real
linear scalar potential, $S(x)=\pm\frac{\lambda}{c}x.$ The energy values and
wavefunctions reduce to$\ $ \ \ \ \
\begin{equation}
\left.  E_{n}\right\vert _{\mu=0}=\pm\frac{c}{\lambda}\sqrt{\left(
2n+1\right)  \hbar A^{3}},\text{ \ \ \ }n=0,1,2,\cdots.\label{31}%
\end{equation}
and%
\begin{equation}
\left.  \psi_{n}\left(  x\right)  \right\vert _{\mu=0}=\left\vert
N_{n}\right\vert e^{i\frac{n\pi}{2}}e^{-\frac{A}{2\hbar}\left(  x+i\frac
{\eta\left.  E_{n}\right\vert _{\mu=0}}{cA^{2}}\right)  ^{2}}H_{n}\left(
\sqrt{\frac{A}{\hbar}}\left(  x+i\frac{\eta\left.  E_{n}\right\vert _{\mu=0}%
}{cA^{2}}\right)  \right)  ,\label{32}%
\end{equation}
with $\left\vert N_{n}\right\vert $ given by (\ref{29}).

\item Setting $\lambda=\mu\omega$, $\eta=\frac{\mu\xi}{c}$ and considering the
non relativistic $\left(  NR\right)  $ limit, by subtracting the rest energy
$\mu c^{2}$ from the total positive energy and taking the limit $c\rightarrow
\infty,$ one gets%
\[
\lim_{c\rightarrow\infty}\left(  E_{n}-\mu c^{2}\right)  =E_{n}^{NR}\text{ and
}\lim_{c\rightarrow\infty}\psi_{n}\left(  x\right)  =\psi_{n}^{NR}\left(
x\right)  ,
\]
with%
\begin{equation}
E_{n}^{NR}=\left(  n+\frac{1}{2}\right)  \hbar\omega.\label{37}%
\end{equation}
and%
\begin{equation}
\psi_{n}^{NR}\left(  x\right)  =\left(  \frac{\mu\omega}{\hbar\pi}\right)
^{\frac{1}{4}}\left(  \frac{1}{2^{n}n!}\right)  ^{\frac{1}{2}}e^{i\frac{n\pi
}{2}}e^{-\frac{\mu\omega}{2\hbar}\left(  x+i\frac{\xi}{\omega^{2}}\right)
^{2}}H_{n}\left(  \sqrt{\frac{\mu\omega}{\hbar}}\left(  x+i\frac{\xi}%
{\omega^{2}}\right)  \right)  .\label{38}%
\end{equation}

We see that in this limit the energy values are those of an harmonic
oscillator and the effect of the vector potential appears only in the wave
functions. In the absence of the vector potential, $\eta=0$ (or $\xi=0$), the
wavefunctions also reduce to those of the harmonic oscillator%
\begin{equation}
\left.  \psi_{n}^{NR}\left(  x\right)  \right\vert _{\eta=0}=\left(  \frac
{\mu\omega}{\pi\hbar}\right)  ^{\frac{1}{4}}\left(  \frac{1}{2^{n}n!}\right)
^{\frac{1}{2}}e^{-\frac{\mu\omega}{2\hbar}x^{2}}H_{n}\left(  \sqrt{\frac
{\mu\omega}{\hbar}}x\right)  .\label{38-a}%
\end{equation}

Thus, on can say that this model may be seen as the extension of the
one-dimensional non relativistic harmonic oscillator to the relativistic
Klein-Gordon harmonic oscillator. Indeed, by setting $\eta=0,$ $\lambda
=\mu\omega$ and taking the non relativistic limit in the Klein-Gordon equation
Eq. (\ref{1}), it can be seen that it reduces to the Schr\"{o}dinger equation
for the harmonic oscillator potential.
\end{itemize}

\subsection{Model with asymptotically bounded mass coupled to a $\mathcal{PT}
$-symmetric hyperbolic vector potential}

In this model, we take the mass distribution and the potential functions in
the forms
\begin{equation}
M\left(  x\right)  =\sqrt{\mu^{2}+\left(  \frac{\lambda}{\alpha c}\right)
^{2}\tanh^{2}\alpha x},\label{39}%
\end{equation}
$\allowbreak$and%
\begin{equation}
V\left(  x\right)  =i\frac{c\eta}{\alpha}\tanh\alpha x,\label{40}%
\end{equation}
where $\mu$ is the rest mass, $\alpha>0$ and $\lambda,\eta$ are real
parameters satisfying (\ref{9-1}), with $\lambda>0.$

Substituting Eqs (\ref{39}) and (\ref{40}) into (\ref{2}) and denoting the
energy by $\mathcal{E}$, the effective $\mathcal{E}$-potential reads%
\begin{equation}
V_{\mathcal{E}}\left(  x\right)  =-\frac{\lambda^{2}+\eta^{2}}{\alpha^{2}%
\cosh^{2}\alpha x}+i\frac{2\eta\mathcal{E}}{\alpha c}\tanh\alpha
x+\frac{\lambda^{2}+\eta^{2}}{\alpha^{2}}+\frac{\mu^{2}c^{4}-\mathcal{E}^{2}%
}{c^{2}},\label{41}%
\end{equation}
that is, for real values of the energy $\mathcal{E}$, a shifted $\mathcal{PT}
$-symmetric potential of Rosen-Morse II type$.$

Choosing the superpotential in the form%
\begin{equation}
W_{\mathcal{E}}\left(  x\right)  =\frac{B}{\alpha}\tanh\alpha x+i\frac
{\eta\mathcal{E}}{cB},\label{42}%
\end{equation}
and using Eq. (\ref{12}), we find that the parameter $B$ and the ground-state
energy $\epsilon_{0}$ are given by%
\begin{equation}
B\left(  B+\hbar\alpha^{2}\right)  =\lambda^{2}+\eta^{2},\label{43}%
\end{equation}
and%
\begin{equation}
\epsilon_{0}=-\left(  \frac{B^{2}}{\alpha^{2}}-\frac{\eta^{2}\mathcal{E}^{2}%
}{c^{2}B^{2}}\right)  +\frac{\lambda^{2}+\eta^{2}}{\alpha^{2}}+\frac{\mu
^{2}c^{4}-\mathcal{E}^{2}}{c^{2}}.\label{45}%
\end{equation}

The unnormalized ground state eigenfunction reads%

\begin{equation}
\Phi_{0}\left(  x\right)  \sim\exp\left(  -\frac{1}{\hbar}\int^{x}%
W_{\mathcal{E}}\left(  y\right)  dy\right)  =e^{-i\frac{\eta\mathcal{E}}{\hbar
cB}x}\left(  \cosh\alpha x\right)  ^{-\frac{B}{\hbar\alpha^{2}}}.\label{45-1}%
\end{equation}

Demanding that $\Psi_{0}\left(  x\right)  $ satisfy the normalization
condition in the sense of Eq. (\ref{I5}), the parameter $B$ must be positive.
Thus, solving Eq.(\ref{43}) with this restriction gives
\begin{equation}
B=\sqrt{\lambda^{2}+\eta^{2}+\frac{\hbar^{2}\alpha^{4}}{4}}-\frac{\hbar
\alpha^{2}}{2}.\label{46}%
\end{equation}
\ \ \ \ \ \ \ 

The supersymmetric partner potentials are constructed as%
\begin{align}
V_{\mathcal{E}}^{\left(  -\right)  }\left(  x\right)   & =W_{\mathcal{E}}%
^{2}\left(  x\right)  -\hbar W_{\mathcal{E}}^{\prime}\left(  x\right)
=-\frac{B\left(  B+\hbar\alpha^{2}\right)  }{\alpha^{2}\cosh^{2}\alpha
x}+i\frac{2\eta\mathcal{E}}{\alpha c}\allowbreak\tanh\alpha x+\frac{B^{2}%
}{\alpha^{2}}-\frac{\eta^{2}\mathcal{E}^{2}}{c^{2}B^{2}},\nonumber\\
& \label{48}\\
V_{\mathcal{E}}^{\left(  +\right)  }\left(  x\right)   & =W^{2}\left(
x\right)  +\hbar W_{\mathcal{E}}^{\prime}\left(  x\right)  =-\frac{B\left(
B-\hbar\alpha^{2}\right)  }{\alpha^{2}\cosh^{2}\alpha x}+i\frac{2\eta
\mathcal{E}}{\alpha c}\allowbreak\tanh\alpha x+\frac{B^{2}}{\alpha^{2}}%
-\frac{\eta^{2}\mathcal{E}^{2}}{c^{2}B^{2}},\nonumber\\
& \label{49}%
\end{align}
which satisfy the shape invariance condition (\ref{a6}), with
\begin{equation}
a_{1}=B,\text{ \ \ }a_{2}=f\left(  a_{1}\right)  =a_{1}-\hbar\alpha
^{2},\label{49-1}%
\end{equation}
and
\begin{equation}
R\left(  a_{1}\right)  =\left(  \frac{a_{1}^{2}}{\alpha^{2}}-\frac{\eta
^{2}\mathcal{E}^{2}}{c^{2}a_{1}^{2}}\right)  -\left(  \frac{a_{2}^{2}}%
{\alpha^{2}}-\frac{\eta^{2}\mathcal{E}^{2}}{c^{2}a_{2}^{2}}\right)
.\label{49-2}%
\end{equation}

By virtue of (\ref{49-1}), one has
\[
a_{k}=\underset{\left(  k-1\right)  \text{ }\mathrm{times}}{\underbrace{f\circ
f\circ\cdots\circ f}}\left(  a_{1}\right)  =a_{1}-\left(  k-1\right)
\hbar\alpha^{2},
\]
such that the energy spectra for bound states of $V_{\mathcal{E}}^{\left(
-\right)  }\left(  x\right)  $ are given by%
\begin{align}
\epsilon_{n}^{\left(  -\right)  }  & =\sum_{k=1}^{n}R\left(  a_{k}\right)
=\left(  \frac{a_{1}^{2}}{\alpha^{2}}-\frac{\eta^{2}\mathcal{E}^{2}}%
{c^{2}a_{1}^{2}}\right)  -\left(  \frac{a_{n+1}^{2}}{\alpha^{2}}-\frac
{\eta^{2}\mathcal{E}^{2}}{c^{2}a_{n+1}^{2}}\right) \nonumber\\
& =\left(  \frac{B^{2}}{\alpha^{2}}-\frac{\eta^{2}\mathcal{E}^{2}}{c^{2}B^{2}%
}\right)  -\left(  \frac{\left(  B-n\hbar\alpha^{2}\right)  ^{2}}{\alpha^{2}%
}-\frac{\eta^{2}\mathcal{E}^{2}}{c^{2}\left(  B-n\hbar\alpha^{2}\right)  ^{2}%
}\right)  .\label{50}%
\end{align}

Using (\ref{21}) and substituting Eqs (\ref{45}) and (\ref{50}) into
(\ref{21}), we get the energy spectra of $V_{\mathcal{E}}(x)$ in the form
\begin{equation}
\epsilon_{n}=\frac{B\left(  B+\hbar\alpha^{2}\right)  }{\alpha^{2}}+\frac
{\mu^{2}c^{4}-\mathcal{E}^{2}}{c^{2}}-\left(  \frac{\left(  B-n\hbar\alpha
^{2}\right)  ^{2}}{\alpha^{2}}-\frac{\eta^{2}\mathcal{E}^{2}}{c^{2}\left(
B-n\hbar\alpha^{2}\right)  ^{2}}\right)  ,\label{51}%
\end{equation}
where $n$ is limited to positive integer numbers satisfying%
\begin{equation}
0\leq n\leq n_{\max}=\left\{  \frac{B}{\hbar\alpha^{2}}\right\}  ,\label{52}%
\end{equation}
and $\left\{  k\right\}  $ denotes the largest integer inferior to $k.$

Thus, the condition $\lambda>0$ is sufficient for the existence of at least
one bound state for the effective potential (real eigenvalues $\epsilon_{n}$
and normalizable corresponding eigenfunctions). However, this will not be
necessarily a sufficient condition for the\ existence of bound states for the
original problem.

Putting $\epsilon_{n}$ $=0$ in Eq. (\ref{51}) and solving it, the allowed
energy values of the original problem, $\mathcal{E}_{n},$ are given by the
following generating formula $\left\{  {}\right\}  $%
\begin{equation}
\mathcal{E}_{n}=\pm\sqrt{\frac{\mu^{2}c^{4}+\frac{c^{2}}{\alpha^{2}}\left(
B\left(  B+\hbar\alpha^{2}\right)  -\left(  B-n\hbar\alpha^{2}\right)
^{2}\right)  }{1-\frac{\eta^{2}}{\left(  B-n\hbar\alpha^{2}\right)  ^{2}}}%
},\label{54}%
\end{equation}
where now allowed values of $n$ must satisfy (\ref{52}) and also are such that
$\mathcal{E}_{n}$ are real. It is easy to see that, while the numerator of the
expression in the square root is always positive if (\ref{52}) is satisfied,
the positivity of the denominator requires the new condition
\begin{equation}
0\leq n\leq\overline{n}_{\max}=\left\{  \frac{B-\eta}{\hbar\alpha^{2}%
}\right\}  ,\label{55}%
\end{equation}
which is more restrictive than (\ref{52}).

Hence, using (\ref{46}) with (\ref{55}) we find that the existence of at least
one bound state for the original problem requires a new constraint on the
parameters $\lambda,\eta$ and $\alpha$, given by%
\begin{equation}
\lambda^{2}>\hbar\alpha^{2}\eta.\label{56}%
\end{equation}

This means that when $\lambda$ increases there is a tendency to increase the
number of bound states, while growth of $\eta$ tends to decrease this number
while magnifying the eigenvalues. In other words, for fixed $\lambda, $ the
number of bound states is maximum for null vector potential ($\eta=0 $) and
then decreases with increasing $\eta.$Thus, the imaginary vector potential
tends to reduce the confinement of the particle that is produced by the
variation of its mass. Indeed, in the case of null vector potential, one has
\begin{equation}
\left.  \mathcal{E}_{n}\right\vert _{\eta=0}=\pm\sqrt{\mu^{2}c^{4}+\frac
{c^{2}}{\alpha^{2}}\left[  \lambda^{2}-\left(  \sqrt{\lambda^{2}+\frac
{\hbar^{2}\alpha^{4}}{4}}-\left(  n+\frac{1}{2}\right)  \hbar\alpha
^{2}\right)  ^{2}\right]  },\label{57}%
\end{equation}
with
\begin{equation}
0\leq n\leq\left.  \overline{n}_{\max}\right\vert _{\eta=0}=\left\{
\sqrt{\frac{\lambda^{2}}{\hbar^{2}\alpha^{4}}+\frac{1}{4}}-\frac{1}%
{2}\right\}  .\label{58}%
\end{equation}

It appears that
\begin{equation}
\left.  \mathcal{E}_{n}\right\vert _{\eta=0}\leq\mathcal{E}_{n},\label{58-1}%
\end{equation}
\qquad and
\begin{equation}
\left.  \overline{n}_{\max}\right\vert _{\eta=0}\geq\overline{n}_{\max
}.\label{58-2}%
\end{equation}

To obtain the wavefunctions $\Psi_{n}\left(  x\right)  $ of the original
problem, $\left(  \Psi_{n}\left(  x\right)  =\left.  \Phi_{n}\left(  x\right)
\right\vert _{\epsilon_{n}\left(  E\right)  =0}\right)  $, we proceed as in
the previous model. In this case, we are led to solve the following equation%
\begin{equation}
\left(  \hbar^{2}\frac{d^{2}}{dx^{2}}+\frac{B\left(  B+\hbar\alpha^{2}\right)
}{\alpha^{2}\cosh^{2}\alpha x}-i\frac{2\eta\mathcal{E}_{n}}{\alpha c}%
\tanh\alpha x+\frac{\eta^{2}\mathcal{E}_{n}^{2}}{c^{2}\left(  B-n\hbar
\alpha^{2}\right)  ^{2}}-\frac{\left(  B-n\hbar\alpha^{2}\right)  ^{2}}%
{\alpha^{2}}\right)  \Psi_{n}\left(  x\right)  =0.\label{59}%
\end{equation}

By the point transformation, defined by
\begin{equation}
z=\tanh\left(  \alpha x\right)  ;\text{ }z\in\left]  -1,1\right[  \text{ and
}\Psi_{n}\left(  x\right)  =\left(  1-z\right)  ^{\frac{a_{n}}{2}}\left(
1+z\right)  ^{\frac{b_{n}}{2}}\phi_{n}\left(  z\right)  ,\text{ }\label{60}%
\end{equation}
with
\begin{subequations}
\begin{align}
a_{n}  & =\frac{B-n\hbar\alpha^{2}}{\hbar\alpha^{2}}+i\frac{\eta
\mathcal{E}_{n}}{\alpha\hbar c\left(  B-n\hbar\alpha^{2}\right)
},\label{61-a}\\
b_{n}  & =\frac{B-n\hbar\alpha^{2}}{\hbar\alpha^{2}}-i\frac{\eta
\mathcal{E}_{n}}{\alpha\hbar c\left(  B-n\hbar\alpha^{2}\right)  }=a_{n}%
^{\ast},\label{61-b}%
\end{align}
it is straightforward to show that the new function $\phi_{n}\left(  z\right)
$ satisfies the differential equation of Jacobi polynomials,
\end{subequations}
\begin{equation}
\left(  1-z^{2}\right)  \frac{d^{2}\phi_{n}\left(  z\right)  }{dz^{2}}+\left[
b_{n}-a_{n}-\left(  a_{n}+b_{n}+2\right)  z\right]  \frac{d\phi_{n}\left(
z\right)  }{dz}+n\left(  n+a_{n}+b_{n}+1\right)  \phi_{n}\left(  z\right)
=0.\label{62}%
\end{equation}

Knowing that $a_{n}$ and $b_{n}$ are the complex conjugates of each other and
taking account of the following symmetry relation of Jacobi polynomials
\cite{Nikiforov}%
\begin{equation}
P_{n}^{\left(  a_{n},b_{n}\right)  }\left(  -z\right)  =\left(  -1\right)
^{n}P_{n}^{\left(  b_{n},a_{n}\right)  }\left(  z\right)  ,\label{63}%
\end{equation}
the wavefunctions $\Psi_{n}\left(  x\right)  $ may be put in a $\mathcal{PT}%
$-symmetric form as follows%
\begin{equation}
\Psi_{n}\left(  x\right)  =\left\vert \mathcal{N}_{n}\right\vert
e^{i\frac{n\pi}{2}}\left(  1-\tanh\alpha x\right)  ^{\frac{a_{n}}{2}}\left(
1+\tanh\alpha x\right)  ^{\frac{b_{n}}{2}}P_{n}^{\left(  a_{n},b_{n}\right)
}\left(  \tanh\alpha x\right)  ,\label{64}%
\end{equation}
where the normalization constant $\left\vert \mathcal{N}_{n}\right\vert $ is
given by%
\begin{equation}
\left\vert \mathcal{N}_{n}\right\vert =\sqrt{\frac{2\alpha n!a_{n}b_{n}%
\Gamma\left(  a_{n}+b_{n}+n+1\right)  }{2^{a_{n}+b_{n}}\left(  a_{n}%
+b_{n}\right)  \Gamma\left(  a_{n}+n+1\right)  \Gamma\left(  b_{n}+n+1\right)
}}=\frac{\left\vert a_{n}\right\vert \sqrt{\alpha n!\Gamma\left(
2\operatorname{Re}a_{n}+n+1\right)  }}{2^{\operatorname{Re}a_{n}}%
\sqrt{\operatorname{Re}a_{n}}\left\vert \Gamma\left(  a_{n}+n+1\right)
\right\vert }.\label{65}%
\end{equation}

\subsubsection{Special case}

\begin{itemize}
\item Taking the limit $\alpha\rightarrow0$ in Eqs (\ref{39}) and (\ref{40}),
the mass distribution and the vector potential coincide exactly with those of
the first model, given \ respectively by (\ref{8}) and (\ref{9}). We will see
that, (\ref{54}), (\ref{64}) and (\ref{65}) reduce also to (\ref{23}),
(\ref{27}) and (\ref{29}), respectively. Indeed, one has%
\begin{equation}
\frac{1}{\alpha^{2}}\left(  B\left(  B+\hbar\alpha^{2}\right)  -\left(
B-n\hbar\alpha^{2}\right)  ^{2}\right)  =B\left(  2n+1\right)  \hbar
-n^{2}\hbar^{2}\alpha^{2}\underset{\alpha\rightarrow0}{\longrightarrow}%
\hbar\sqrt{\lambda^{2}+\eta^{2}}\left(  2n+1\right)  ,\label{66}%
\end{equation}
and%
\begin{equation}
1-\frac{\eta^{2}}{\left(  B-n\hbar\alpha^{2}\right)  ^{2}}\underset{\alpha
\rightarrow0}{\longrightarrow}\frac{\lambda^{2}}{\lambda^{2}+\eta^{2}%
},\label{67}%
\end{equation}
such that
\begin{equation}
\mathcal{E}_{n}\underset{\alpha\rightarrow0}{\longrightarrow}E_{n}=\pm
\frac{\sqrt{\lambda^{2}+\eta^{2}}}{\lambda}\sqrt{\mu^{2}c^{4}+\left(
2n+1\right)  \hbar c^{2}\sqrt{\left(  \lambda^{2}+\eta^{2}\right)  }%
},\label{68}%
\end{equation}
which is exactly the relation (\ref{23}).\ The number of energy levels is now
unlimited, i.e. $n=0,1,2,...$, as it can be verified by taking the limit
$\alpha\rightarrow0$ in (\ref{55}). In addition, in the limit $\alpha
\rightarrow0,$ the constraint (\ref{56}) reduces to $\lambda>0$ as it should be.

As regards the wave functions, keeping only leading terms in the limit
$\alpha\rightarrow0$ leads to%
\begin{equation}
\operatorname{Re}a_{n}\underset{\alpha\rightarrow0}{\longrightarrow}%
\frac{\sqrt{\lambda^{2}+\eta^{2}}}{\hbar\alpha^{2}}-\left(  n+\frac{1}%
{2}\right)  \text{ and }\operatorname{Im}a_{n}\underset{\alpha\rightarrow
0}{\longrightarrow}\frac{\eta E_{n}}{\hbar c\alpha\sqrt{\lambda^{2}+\eta^{2}}%
},\label{69}%
\end{equation}
such that a straightforward calculation gives%
\begin{equation}
\left(  1-\tanh\alpha x\right)  ^{\frac{a_{n}}{2}}\left(  1+\tanh\alpha
x\right)  ^{\frac{a_{n}^{\ast}}{2}}\underset{\alpha\rightarrow
0}{\longrightarrow}\exp\left(  -\frac{\sqrt{\lambda^{2}+\eta^{2}}}{2\hbar
}x^{2}-i\frac{\eta\bar{E}_{n}}{\hbar c\sqrt{\lambda^{2}+\eta^{2}}}x\right)
,\label{69a}%
\end{equation}
and (See definitions (8.960.1, page 999) and (8.950.1, page 996) in Ref.
\cite{Gradshteyn})%
\begin{equation}
P_{n}^{\left(  a_{n},b_{n}\right)  }\left(  \tanh\alpha x\right)
\underset{\alpha\rightarrow0}{\longrightarrow}\frac{1}{2^{n}n!}\left(
\frac{\sqrt{\lambda^{2}+\eta^{2}}}{\alpha^{2}\hbar}\right)  ^{\frac{n}{2}%
}H_{n}\left(  \left(  \frac{\sqrt{\lambda^{2}+\eta^{2}}}{\hbar}\right)
^{\frac{1}{2}}\left(  x+i\frac{\eta E_{n}}{c\left(  \lambda^{2}+\eta
^{2}\right)  }\right)  \right)  .\label{70}%
\end{equation}

\end{itemize}

Otherwise, using Stirling formula%
\begin{equation}
\Gamma\left(  X\right)  =\sqrt{2\pi}X^{X-\frac{1}{2}}e^{-X},\label{71}%
\end{equation}
that is valid for large $X,$ a straightforward calculation leads to%
\begin{equation}
\Gamma\left(  2\operatorname{Re}a_{n}+n+1\right)  \underset{\alpha
\rightarrow0}{\longrightarrow}\sqrt{2\pi}\left(  2\frac{\sqrt{\lambda^{2}%
+\eta^{2}}}{\alpha^{2}\hbar}\right)  ^{\frac{2\sqrt{\lambda^{2}+\eta^{2}}%
}{\alpha^{2}\hbar}-n-\frac{1}{2}}e^{-2\frac{\sqrt{\lambda^{2}+\eta^{2}}%
}{\alpha^{2}\hbar}},\label{72}%
\end{equation}
and%
\begin{equation}
\left\vert \Gamma\left(  a_{n}+n+1\right)  \right\vert \underset{\alpha
\rightarrow0}{\longrightarrow}\sqrt{2\pi}\left(  \frac{\sqrt{\lambda^{2}%
+\eta^{2}}}{\alpha^{2}\hbar}\right)  ^{\left(  \frac{\sqrt{\lambda^{2}%
+\eta^{2}}}{\alpha^{2}\hbar}\right)  }\exp\left(  -\frac{\eta^{2}E_{n}^{2}%
}{2\hbar c^{2}\left(  \lambda^{2}+\eta^{2}\right)  ^{\frac{3}{2}}}\right)
e^{-\frac{\sqrt{\lambda^{2}+\eta^{2}}}{\hbar\alpha^{2}}}.\label{73}%
\end{equation}

Using (\ref{69}), (\ref{72}) and (\ref{73}), we easily verify that, in the
leading order of $\alpha,$ the normalization constant $\left\vert
\mathcal{N}_{n}\right\vert $ reduces to%
\begin{equation}
\left\vert \mathcal{N}_{n}\right\vert \underset{\alpha\rightarrow
0}{\longrightarrow}\left\vert \overline{\mathcal{N}}_{n}\right\vert
=\sqrt{2^{n}n!}\left(  \frac{\sqrt{\lambda^{2}+\eta^{2}}}{\pi\hbar}\right)
^{\frac{1}{4}}\left(  \frac{\sqrt{\lambda^{2}+\eta^{2}}}{\alpha^{2}\hbar
}\right)  ^{-\frac{n}{2}}\exp\left(  \frac{\eta^{2}E_{n}^{2}}{2\hbar
c^{2}\left(  \lambda^{2}+\eta^{2}\right)  ^{\frac{3}{2}}}\right)  .\label{74}%
\end{equation}

Finally, substitution of (\ref{69a}), (\ref{70}) and (\ref{74}) into
(\ref{64}) leads to%
\begin{align}
\Psi_{n}\left(  x\right)  \underset{\alpha\rightarrow0}{\longrightarrow}%
\psi_{n}\left(  x\right)   & =\frac{e^{i\frac{n\pi}{2}}}{\sqrt{2^{n}n!}%
}\left(  \frac{\sqrt{\lambda^{2}+\eta^{2}}}{\pi\hbar}\right)  ^{\frac{1}{4}%
}e^{-\frac{\sqrt{\lambda^{2}+\eta^{2}}}{2\hbar}\left(  x+i\frac{\eta E_{n}%
}{c\left(  \lambda^{2}+\eta^{2}\right)  }\right)  ^{2}}\nonumber\\
& \times H_{n}\left(  \sqrt{\frac{\sqrt{\lambda^{2}+\eta^{2}}}{\hbar}}\left(
x+i\frac{\eta E_{n}}{c\left(  \lambda^{2}+\eta^{2}\right)  }\right)  \right)
,\label{75}%
\end{align}
that coincide exactly with the wave function of the first model (relations
(\ref{27}) and (\ref{29}).

\section{Conclusion}

In this paper, we have discussed bound state solutions of the $\left(
1+1\right)  $-dimensional stationary Klein-Gordon equation with
position-dependent mass and $\mathcal{PT}$-symmetric vector and scalar
potentials by the approach of supersymmetric quantum mechanics. We have shown
that, for better use of SUSYQM, the problem can be mapped into a constant mass
Schr\"{o}dinger equation with energy-dependent effective potential. This
method is applied to solve exactly two models with null scalar potentials and
suitable couples of mass distribution and $\mathcal{PT} $-symmetric vector
potential, that, interestingly, coincide in a limiting case.

In the first model, the vector potential is chosen as a ${\mathcal{PT}}$-symmetric linear function of the position, and the mass distribution is the
square root of a quadratic form. The problem leads to solve Schr\"{o}dinger
equation with quadratic energy-dependent ${\mathcal{PT}}$-symmetric
potential.\ The bound-state energies are exactly obtained by SUSYQM and the
wavefunctions are easily deduced.

In the second model, the ${\mathcal{PT}}$-symmetric vector potential is
chosen as a hyperbolic tangent function, and the mass distribution is the
square root of a quadratic form of a hyperbolic tangent function. The problem
is then reduced to solve Schr\"{o}dinger equation with an energy-dependent 
${\mathcal{PT}}$-symmetric potential of Rosen-Morse II type. Again,
SUSYQM approach has been applied successfully to obtain exactly the
bound-state energies and to deduce the corresponding wavefunctions. In
particular, we have discussed the constraints that must be satisfied by the
parameters of the problem in order to obtain physical results. Furthermore, we
have discussed some special cases of the two models and shown that they
coincide in a limiting case.

\end{document}